\documentstyle[11pt,aaspp4]{article}

\received{March 11, 1996}
\accepted{May XX, 1996}
\journalid{}{}

\newcommand{\etal}{et al.}
\newcommand{\eg}{e.g.}

\newcommand{\hone}{\mbox{$h^{-1}$}}
\newcommand{\htwo}{\mbox{$h^{-2}$}}
\newcommand{\perone}{\mbox{$^{-1}$}}
\newcommand{\pertwo}{\mbox{$^{-2}$}}
\newcommand{\lam}{\mbox{$\lambda$}}
\newcommand{\mgii}{\ion{Mg}{2}}
\newcommand{\oi}{[\ion{O}{1}]}

\newcommand{\hei}{\ion{He}{1}}
\newcommand{\ha}{H$\alpha$}

\newcommand{\nii}{[\ion{N}{2}]}
\newcommand{\sii}{[\ion{S}{2}]}
\newcommand{\HST}{{\sl HST}}
\newcommand{\IRAS}{{\sl IRAS}}
\newcommand{\ex}[2]{\mbox{#1$\times$10$^{#2}$}}

\begin{document}


\title{IRAS FSC 15307+3252:\\Gravitationally Lensed Seyfert or
Cannibal Elliptical at $z$ = 0.93?\altaffilmark{1}}

\author{Michael C. Liu and James R. Graham\altaffilmark{2}}
\affil{Department of Astronomy, University of California,
Berkeley, CA 94720\\email: {\tt [mliu,jrg]@astro.berkeley.edu}}

\and

\author{Gillian S. Wright}
\affil{Joint Astronomy Centre, 660 N. A'oh$\bar{\rm o}$k$\bar{\rm
u}$ Place, University Park, Hilo, HI 96720\\email: {\tt
gsw@jach.hawaii.edu}}

\altaffiltext{1}{Based in part on observations obtained at the W.M. Keck
Observatory, which is operated jointly by the University of California
and the California Institute of Technology.}
\altaffiltext{2}{Alfred P. Sloan Fellow.}

\begin{center}
To appear in the October 20, 1996 issue of the {\em Astrophysical Journal}
\end{center}

\begin{abstract}
We present the highest spatial and spectral resolution
near-infrared data to date of the $\sim~10^{13}~\htwo~L_{\sun}$
{\sl IRAS} source FSC 15307+3252 at $z = 0.93$, apparently the most
luminous galaxy in the known Universe.  Deep $K$-band (2.2
\micron) images taken in 0\farcs4 seeing at the W.\ M.\ Keck
Telescope reveal three components: (A) a bright elliptical source
with a compact nucleus, (B) a resolved circular companion separated
from component A by 2\farcs0 (8$h^{-1}$~kpc for $q_{0} = 0.5$), and
(C) a faint irregular component 1\farcs7 from A.  The surface
brightness profile of F15307-A is well-characterized by a
de~Vaucouleurs $r^{1/4}$ law with $r_{e} = 1\farcs4 \pm 0\farcs2$
(6\hone~kpc), a size comparable to local giant ellipticals. The
nucleus of component A is stellar in appearance with extended
structure, possibly a second nucleus $\sim$~0\farcs5 away.  Our
1.1--1.4~\micron\ spectrum of the F15307 system with a resolution
of 330 km~s\perone\ shows strong emission lines of \oi\
\lam\lam6300, 6364; blended \ha~+~\nii\ \lam\lam6548, 6583; and
\sii\
\lam\lam6716, 6731.  The $\sim$~900~km~s\perone\ width of the
forbidden lines and the relative strengths of the emission lines
are characteristic of Seyfert 2 galaxies.  The \ha\ line also has a
broad (1900 km~s\perone) component.

In light of the recent discovery that FSC 10214+4724, previously
the most luminous known galaxy, is a gravitationally-lensed system,
we explore the possibility that F15307 is also lensed.
Quantitative arguments are inconclusive, but aspects of F15307's
morphology do suggest lensing; the system bears a strong
resemblance to quadruple-image gravitational lenses. On the other
hand, given the $r^{1/4}$ profile, the close companions, and the
active nucleus, F15307 may in fact be a giant elliptical galaxy
caught in the act of galactic cannibalism, a scenario which could
also account for its unparalleled luminosity.
\end{abstract}


\section{Introduction}

During the course of an \IRAS\ color-selected survey for extremely
luminous IR-bright galaxies, Cutri \etal\ (1994) identified FSC
15307+3252 as an $\sim 1.0 \times 10^{13}\ h^{-2} L_{\sun}$ galaxy
at a redshift of 0.926 ($q_{0} = 0.5$, $H_{0} = 100h$
km~s\perone~Mpc\perone). They found its restframe UV/blue optical
spectrum resembles a Seyfert 2 galaxy; Soifer \etal\ (1994) found a
similar result for the restframe red optical emission lines though
they had insufficient spectral resolution to measure linewidths.
Hines \etal\ (1995) have found broad \ion{Mg}{2} $\lambda$2798
emission and a power-law continuum in polarized light, leading them
to argue the system contains a buried quasar.  IR imaging by Soifer
\etal\ (1994) in 1\arcsec\ seeing shows the system is composed of a
bright extended source with one or two close companions.

The one known galaxy believed to be more luminous than F15307, the
$\sim 5
\times 10^{13}\ h\pertwo\ L_{\sun}$ source FSC 10214+4724 at $z =
2.286$, is now known to be the first example of a
gravitationally-lensed Seyfert 2 galaxy.  Its $I$-band flux is
magnified by a factor of 100 (\cite{eis96}), and the $K$-band
magnification is around 10--20 (\cite{gra95,bro95}).  In hindsight,
this result is not surprising.  Many known lensed systems are $z
\gtrsim 1$ quasars. In the local universe, the space density of 
luminous ($L \gtrsim 10^{11} L_{\sun}$) IR galaxies exceeds that of
quasars (\cite{soi86}); if this fact holds at higher redshifts, one
naturally expects gravitational lensing of IR-bright
galaxies. \IRAS\ sources at $z \gtrsim 0.5$ are prime suspects for
this phenomenon since the associated magnification would allow
these objects to be detected at significant redshifts.  Statistical
estimates support this line of reasoning (\cite{bro95,tre95}).
However, while lensed quasars are relatively easy to identify since
lensing of point sources produces distinctive sets of multiple
images, lensed extended sources, like IR-bright galaxies, are more
difficult to recognize for two reasons: (1) their total
magnification is less, and (2) they form extended images which
require high angular resolution to resolve, \eg, sub-arcsecond
interferometric imaging is needed to identify lensed high redshift
radio lobes (\eg, \cite{bla92}).

Therefore, high resolution imaging of F15307 is necessary in order
to determine whether its extraordinary luminosity arises from
gravitational lensing or intrinsic phenomena such as massive
starbursts and/or an active nucleus.  If F15307 and other high-$z$
\IRAS\ sources are gravitationally lensed like F10214, this
discovery will have applications beyond understanding the nature of
these objects.  Lensed extended sources are useful tools for
probing the mass distribution of the lensing galaxies since they
offer more lines of sight through the lens than lensed point
sources (\eg,
\cite{koc91}).  Imaging provides a more effective tool to search
for lensing than spectroscopy, since the latter typically requires
high S/N to search for discrepant redshifts from continuum
features.  Targets found with morphologies suggestive of lensing can then
be spectroscopically examined to determine a redshift for the
foreground lens.  Ultimately, statistics of these lensed systems
will quantify the magnification bias afflicting the high end of the
\IRAS\ luminosity function.

In this paper we present 0\farcs4 resolution $K$-band
imaging and moderate resolution ($\lambda/\Delta\lambda$ = 990)
near-IR (1.1--1.4 \micron) spectroscopy of FSC 15307+3252.  Our
imaging data identify three components all within 2\arcsec.  The
near-IR (restframe optical) spectrum displays emission lines of
\oi, \ha~+~\nii, and \sii\ which resemble a Seyfert~2-type spectrum.  The
morphology of the system is similar to quadruple-image
gravitational lenses, though at the limit of our resolution, we
cannot discount the possibility the system is involved in a close
interaction and/or merger of 2--3 separate components; in particular,
the brightest component appears to be a large elliptical galaxy in the
process of assembling.  Throughout the paper we assume $q_{0} =
0.5$ and $H_{0} = 100h$ km~s\perone~Mpc\perone.  With this choice
of cosmology, 1\farcs0 corresponds to 4.2\hone\ kpc at a redshift
of 0.926.


\section{Observations and Data Reduction}

\subsection{Imaging}

We observed F15307 on 1995 May 23 (UT) using the facility near-IR
camera (\cite{mat94}) of the 10-meter W.\ M.\ Keck Telescope
located on Mauna Kea, Hawaii.  The camera employs a Santa Barbara
Research Corporation 256 $\times$ 256 InSb array and has a plate
scale of 0\farcs15 pixel\perone.  We observed using the standard
$K$ (2.0--2.4 \micron) filter.  We used an integration time of 10 s
per coadd, and after six or ten frames were coadded and saved as an
image, we offset the telescope by a few arcseconds.  The telescope
was stepped using a non-redundant dither pattern.  An off-axis CCD
camera was used to guide the telescope during the observations.
The total integration time was 2100 s.

We constructed a flat field by median averaging the images after
subtraction of a dark frame to remove the bias level; a preliminary
sky subtraction was performed to identify astronomical objects so
as to exclude them from the averaging.  Then for each frame, we
subtracted a local sky frame constructed from the average of prior
and subsequent frames, again excluding any astronomical objects
from the averaging.  We used the bright star $\approx$ 20\arcsec\
north of F15307 to register the reduced frames and then shifted by
integer pixel offsets to assemble a mosaic of the field.  Bad
pixels were identified and masked during the construction of the
mosaic; we avoided interpolating pixel values.  We performed all
the reductions and analysis using Research System Incorporated's
IDL software package (version 4.0.1) unless otherwise noted.

We observed the faint UKIRT standard FS 27 ($K = 13.123~\pm~0.018$;
\cite{cas92}) immediately before F15307 as a flux calibrator.  Our
derived zero point, the magnitude of a star which produces 1 count
for a 1 second integration, is 25.22~$\pm$~0.02 mag.  The sky was
moonless so it was difficult to determine at the time if it was
clear. Photometry of the aforementioned bright star in the
individual frames had a standard deviation of 5\% with a slight
trend toward fewer counts during the course of the observations.
We measure a $K$ magnitude for this star 0.08 mag fainter than
Soifer \etal\ (1994), and taking this offset into account, our
photometry for F15307 in 6\arcsec--9\arcsec\ diameter apertures is
consistent with theirs. Throughout the paper we use our derived
photometry as any systematic errors are of order 5\% or less.

Figure 1 shows our final $K$-band mosaic.  The central 25\arcsec\
$\times$ 25\arcsec\ region achieves 1 $\sigma$ noise of 23.9 mag
arcsec$^{-2}$.  The faintest objects in the image are $K\approx
21.5-22$ mag. The spatial resolution of the mosaic, as determined
from the full width at half maximum (FWHM) of the brightest star in
the field, is 0\farcs4.  For brevity, we will refer to this star as
the ``bright star'' (called ``star A'' by
\cite{soi94}).  Nearly all other objects in the mosaic are
extended.

Table 1 presents relative astrometry and photometry for objects in
the field identified by eye.  With the exception of F15307, the
photometry is measured in a 3\arcsec\ diameter aperture using a
6\arcsec--9\arcsec\ diameter annulus for sky level determination
and adjusted by a small (0.03 mag) aperture correction derived from
the bright star.  The components of F15307 were measured in
2\arcsec\ (component A), 1\farcs5 (B), and 1\arcsec\ (C) diameter
apertures modified by the appropriate aperture corrections; as
we discuss in \S~3.2, the expected contamination from $K$-band line
emission is negligible.  Photometry errors were calculated by
combining in quadrature the noise in the photometry aperture with
errors in the zero point and sky level determination.  The number
and brightnesses of the objects in the F15307 field are consistent
with $K$-band galaxy number counts (\cite{gar93,djo95}). Note that
using the calculations from \S~4.1, components A and B have $K$
magnitudes equivalent to $\sim$ 5 and 1 $L^*$ ellipticals at
$z=0.926$, respectively.

\subsection{Spectroscopy}

We obtained a $J$-band spectrum of F15307 on 1995 July 13 (UT)
using the facility long slit spectrograph CGS4 (\cite{mou90}) of
the United Kingdom Infrared Telescope (UKIRT) located on Mauna Kea,
Hawaii.\footnote{UKIRT is operated by the Royal Observatory,
Edinburgh, on behalf of the UK Science and Engineering Research
Council.}  We employed the 75~lines~mm\perone\ grating in second
order to obtain a nominal resolution ($\lambda/\Delta\lambda$) of
990 (330~km~s\perone) with 0.33~\micron\ wavelength coverage, and
we oriented the slit to within 1\arcdeg\ of north-south.
Conditions were photometric.  The spectrograph camera images one
pixel across the 1\farcs5 slit so in order to fully sample the
instrument profile the detector was stepped 1/2 pixel in the
spectral direction after a 30 second on-chip exposure time.  The
source was nodded between two positions on the slit every minute,
after each fully sampled spectrum was obtained. The observation
consisted of 20 such nodded pairs for total time exposure of 40
minutes.  Observations of the blackbody and krypton lamp in the
CGS4 calibration unit were obtained for flat field and wavelength
calibration.  The nearby G3V star HR~5728 was observed with the
same nodding technique before and after observing F15307 as an
atmospheric standard, and a spectrum of HD~136754 (\cite{eli82})
was obtained for flux calibration.

Data reduction was carried out using the Figaro data reduction
package (\cite{sho93}). Each individual spectrum was flat fielded,
the nodded data subtracted in pairs, and the pairs coadded.
Wavelength calibration from the krypton lamp was accurate to
0.0005~\micron, as determined by our measurement of Paschen~$\beta$
absorption in HD~136754 at 1.2827~\micron.  We used the method of
Horne (1986) to extract the spectra from the two slit positions and
then summed the extractions.  The spectra of the two standard stars
were reduced and extracted in the same fashion as the galaxy
spectrum.  Atmospheric absorption in the spectrum of F15307 was
corrected by using the spectrum of HR~5728, which was corrected to
the same mean airmass and divided by a 5770~K blackbody.  Flux
calibration was derived from the spectrum of HD~136754, assuming 
a $J$ magnitude of 7.135.

Figure 2 presents our reduced spectrum.
The spectrum of the F15307 system possesses emission lines of
\oi~$\lambda\lambda$6300,~6364;
\ha~+~\nii~$\lambda\lambda$6548,~6583;  and 
\sii~$\lambda\lambda$6716,~6731.  The possible 
narrow feature at 1.165~\micron\ lies $\sim$ 3000 km~s\perone\
blueward of the expected position of [\ion{Fe}{7}]~$\lambda$6087;
curiously, the very broad ($\sim$ 10$^4$ km~s\perone) polarized
\mgii\
\lam2798 line detected by Hines \etal\ (1995) is blueshifted this
same amount from the unpolarized narrow \mgii\ component.  If the
[\ion{Fe}{7}] identification is correct the feature is unusually
strong compared to other narrow-lined AGN (\cite{ost89}), but this
line can be strong in highly photoionized regions such as the radio
galaxy PKS~2152--69 (\cite{tad88}).


\section{Data Analysis and Results}

\subsection{$K$-band Morphology}

Figure 3 shows the structure of F15307 and the designations we will
use for the three prominent components.  The brightest source,
component A, is elliptical with a position angle (PA) of $\approx$
40\arcdeg\ and a major to minor axis ratio of $\approx$ 1.2.  The
PA is the angle between north and the major axis of component A,
measured east from north.
Component B, the next brightest source, lies 2.0\arcsec\
(8\hone~kpc) to the southeast of component A.  It is resolved into an
extended source and is roughly circular.  The faintest source,
component C, is 1\farcs7 east of component A and far more irregular
in appearance than the other two sources.  Low-level diffuse
emission surrounds the entire system, with a total extent of about
6\arcsec\ (25\hone~kpc) and roughly an elliptical shape with a
different PA than component A. Based on $K$ band number counts, it
is unlikely the three components are a chance superposition of
objects at different redshifts (\cite{soi94}).

\subsubsection{F15307-A: An Elliptical Galaxy with 
Nuclear Structure}

Given the excellent angular resolution of the image, we are able to
study quantitatively the morphology of component A.  We extracted
the surface brightness (SB) profile of this component using
elliptical apertures with the aforementioned PA and axial ratio
spaced by the seeing FWHM.  A 6\arcsec--9\arcsec\ radius elliptical
annulus was used to determine the sky level.  We included pixels
only from the northwestern half of F15307-A (PA from 215\arcdeg\ to
35\arcdeg) to avoid contamination from the other two components and
computed errors by summing in quadrature the standard error in each
annulus with the (essentially negligible) sky error determined from
the scatter in the sky annulus pixels. We then fitted the profile
from 1\arcsec--3\arcsec\ in semi-major axis with de~Vaucouleurs
$r^{1/4}$ and exponential disk profiles (\eg, \cite{mih81}) using a
standard non-linear least-squares 
algorithm; we avoided the central 1\arcsec\ to reduce seeing
effects and contamination by any unresolved nuclear component. The
de~Vaucouleurs profile provided a very good fit while the
exponential profile did not, as judged both by eye and by the
reduced chi-square values ($\tilde{\chi}^2 $= 0.4 versus 4.7).  The
isophotes of component A rotate slightly over the fitting range,
but this fact should not strongly affect the fitting.  Figure 4
displays component A's SB profile along with the de~Vaucouleurs and
exponential fits.

To quantify systematic effects from seeing, we repeated the fitting
process on artificial models of galaxies with de~Vaucouleurs and
exponential profiles over a range of SBs and scale lengths.  We
convolved the artificial galaxies with the bright star as a
representation of the point spread function (PSF), subpixellating
the model and psf before convolution and rebinning afterwards.  We
then added Gaussian noise equal to the amount in the reduced image
of F15307 and applied the fitting algorithm to the artificial
galaxies.  For a given model galaxy, we added noise and performed the
fit many times.  Our tests confirm we can effectively distinguish
between de~Vaucouleurs and exponential models over the relevant
range of parameters and that the formal statistical errors derived
from the fit are reasonable.  Our tests also suggest we have
systematically underestimated $\mu_e$ by $\simeq$ 0.5
mag~arcsec$^{-2}$ and $r_e$ by $\simeq 20\%$.  For F15307-A we
derive final values for the $K$-band de~Vaucouleurs profile of
$\mu_{e} = 20.1 \pm 0.3$ mag~arcsec\pertwo\ and $r_{e} = 1\farcs4
\pm 0\farcs2$ (6\hone~kpc).

Figure 5 shows our image of F15307 after subtraction of a
seeing-convolved de~Vaucouleurs profile juxtaposed against the PSF.
The residuals show a bright compact source at the center F15307-A
with a sub-arcsecond extension to the southwest (PA $\sim$
220\arcdeg).  The flux in a 1\farcs5 diameter aperture centered on
the core is 17.6 mag, about 1/3 of the total flux in this region.
This source likely does not significantly contaminate our
de~Vaucouleurs fits; applying our SB fitting procedure on the PSF
shows the PSF wings have a much smaller scale length than component
A.  Our analysis in \S~3.1.2 verifies the reality of this nuclear
structure and hints that the extension seen in the residual image is
a second compact source $\sim$~0\farcs5 (2\hone~kpc) from the
central nucleus and $\sim$~10 times fainter.  This 
structure is also present in images deconvolved with the
Lucy-Richardson algorithm (\cite{luc74,ric72}) as implemented in IRAF
(version 2.10.3).\footnote{IRAF (Image Reduction and Analysis
Facility) is distributed by the National Optical Astronomy
Observatories, which are operated by the Association of
Universities for Research in Astronomy, Inc., under cooperative
agreement with the National Science Foundation.}

\subsubsection{The Companions: F15307-B and F15307-C}

To study components B and C better, we removed the emission
associated with component A assuming elliptical symmetry.  In
addition to subtracting a de~Vaucouleurs profile (\S~3.1.1), we
removed component A using two other methods: (1) by subtracting its
SB profile directly from the image and (2) by subtracting a
180\arcdeg\ rotated version of the NW half from the SE half. All
methods produced comparable results to the image shown in Figure 5.
The relative astrometry of the three components in Table 1 uses the
averages of measurements taken from the subtracted image and the
original image.  The error in the relative positions of A and B is
$\sim$ 0\farcs02 and between A and C is $\sim$~0\farcs04 based on
the results from the different methods.

Component B is circular and clearly extended with a FWHM of
0\farcs5.  Its $K$ magnitudes in 1\farcs5 and 2\farcs5 diameter
apertures with a point source-derived aperture correction are
18.6~$\pm$~0.1 and 18.2~$\pm$~0.1, respectively.  There is faint
emission along the axis connecting A and B.

Component C is noticeably more diffuse than either A or B.  It may
not be a distinct component at all but rather a tidal feature,
though it would then have to be intrinsically bright to overcome
cosmological surface brightness dimming.  Its inner ($\sim
0\farcs5$ diameter) isophotes are extended roughly north-south
while the $\sim$1\arcsec\ diameter isophotes are approximately
perpendicular to the axis connecting A and B and extended
preferentially to the NE. Aperture-corrected photometry in a
1\arcsec\ diameter aperture gives $K \approx 20.1 \pm 0.1$~mag.

\subsection{Emission Line Spectrum}

We measured each set of close lines (\oi, \ha~+~\nii, \sii)
independently, approximating each line as a Gaussian.  FWHMs for
the lines of each doublet were constrained to be the same, and the
FWHM was allowed to vary between doublets.  The line ratios of the
\oi\ and \nii\ doublets were fixed.  We varied the \sii\ line ratio
from the low to high density limit; the high density limit fit
slightly better though it produced a greater redshift discrepancy
between \oi\ and \sii.  Our measured \oi\ and \sii\ FWHMs are
$900~\pm~300$ km~s\perone\ and $900~\pm~600$ km~s\perone,
respectively, comparable to those of the restframe blue/UV
forbidden lines (\cite{cut94}).

We deblended \ha~+\nii\ with three methods.  First, we constrained the
\nii\ linewidths to be the same as \oi\ and \sii;
\nii\ has a critical density in between the two so this assumption
is reasonable (\cite{fil84,fil85,der86}).  In the second method,
both the \ha\ and \nii\ linewidths were allowed to vary.  In the
third method, we constrained the FWHM of \nii\ and \ha\ to be the
same as
\oi\ and \sii\ and added a broad component of \ha, attempting to
account for narrow \ha\ emission from starbursts and HII regions
and broader emission from an AGN.  In all the methods, the
redshifts of the \ha\ and \nii\ lines were free parameters. All
methods find comparable redshifts, FWHMs, and equivalent widths for
the lines as well as a significant broad (1910~$\pm$~50
km~s\perone\ FWHM) component to the \ha\ line.  We conservatively
constrain any very broad \ha\ component to have a restframe
equivalent width $\lesssim$ 0.10
\micron, assuming it has the same $\sim~10^4$ km~s\perone\ FWHM as
the polarized broad \mgii\ \lam2798 line (\cite{hin95}).

Table 2 provides the extracted physical parameters.  The errors
are from the formal fitting errors.  We list results from the
high-density fit to \sii\ and the deblending of
\ha~+~\nii\ which incorporates a narrow and broad \ha\ component.  The total
observed equivalent width of the lines is 0.10 $\pm$ 0.01~\micron,
which accounts for $\approx$ 40\% of the spectrum's $J$-band flux.
Using the \ha\ flux, we can predict the emission line contamination
to the $K$-band observations is negligible.  The strongest relevant
line is \hei\ \lam10830 (\cite{rud89,ost90,rud93}).  Using typical
flux ratios of \hei\ to \ha\ observed by Rudy \etal\ (1989) in
Seyfert 2 galaxies, the $K$-band line contamination to the
integrated F15307 flux is expected to be $\lesssim$ 5\% and
$\lesssim$~10\% if we consider only the inner 2\arcsec\ of F15307-A
and assume it is the lone source of line emission.

A variety of diagnostics show the spectrum is Seyfert 2-type.  The
FWHM of the lines is consistent with such a spectrum, though both
the permitted and forbidden linewidths are somewhat greater than
typical of Seyfert~2's (\cite{ost89,ost91}). We also classified the
spectrum using the excitation diagrams of Veilleux \& Osterbrock
(1987), supplemented with data from Ho, Filippenko, \& Sargent
(1993).  We took the [\ion{O}{3}] \lam5007 flux from Soifer
\etal\ (1994) and assumed \ha/H$\beta = 2.9$, appropriate for pure
recombination with $T = 10^4$~K (\cite{ost89}).  The
location of F15307 in these diagrams is consistent with Seyfert 2
excitation.  Though much of the width of the forbidden lines might
be due to $\sim$~300~km~s\perone\ relative orbital motion between
two of the components encompassed in our slit, such an aperture
effect should not affect the Seyfert 2 classification from the
excitation diagrams (\cite{ho93a}) nor the presence of the broad
\ha\ component.

Our restframe equivalent widths for
\oi\ are larger and \ha~+~\nii\ and \sii\ are smaller than Soifer
\etal\ (1994).  Since our slit was 1\farcs5 and theirs was 0\farcs6,
the differences may indicate that the \oi\ emission region is
considerably extended compared to the source(s) of the other lines.
Local luminous \IRAS\ galaxies do show extended line emission with
excitation gradients (\cite{arm90,feh94,vei95}).


\section{Discussion}

\subsection{An Example of Gravitational Lensing?}

In light of the discovery that F10214 is a gravitationally lensed
system, it is natural to ask if the apparent extraordinary
luminosity of F15307 is also due to lensing.  We can use the
morphological information in our high angular resolution data to
explore the lensing scenario in a quantitative fashion.  To review
\S~3.1, F15307 can be decomposed into three components: (A) a large
elliptical source with a bright star-like core and elongated
nuclear structure; (B) a resolved circular companion separated
2\farcs0 from A; and (C) a diffuse irregular companion
$\sim$1\farcs7 from A.  The most plausible feature which could be
ascribed to gravitational lensing is the nuclear structure of
component A.  The bright nucleus and close extension or second
nucleus could be due to multiple imaging/stretching of a $z=0.93$
source by a foreground lens.

A natural suspect for the foreground lens is component
B because the nuclear structure of A would then be tangential image
stretching.  To examine this hypothesis, we can calculate the
expected angular separation between the lens and images based on an
estimated lens mass as a function of assumed redshift.  The
Einstein ring radius for an isothermal sphere with a
one-dimensional velocity dispersion $\sigma$ is $$\theta_{E} =
1\farcs40\
\left({{\sigma}\over{220\ {\rm km\ s^{-1}}}}\right)^2\
{{D_{LS}}\over{D_{OS}}}$$ where $D$ is the angular diameter
distance from lens to source (LS) and observer to source (OS),
respectively (\cite{bla92}).  To estimate $\sigma$ for F15307-B, we
start with its $K$ magnitude in a 2\farcs5 diameter aperture
(\S~3.1.2) and apply a small correction from Frogel \etal\ (1978)
to adjust to an aperture of $D_0$, assuming B is an elliptical with
$D_0$~=~30\hone~kpc.  We take $K$-corrections from unevolving
elliptical galaxy spectral templates of Bruzual \& Charlot (1993)
and choose $M_{B}^{*} = -21.3$~mag (\cite{efs88}) to determine
the luminosity of B.  We then use the Faber-Jackson relation, $L
\sim (\sigma_{\parallel})^4$, to convert the luminosity to
line-of-sight stellar velocity dispersion $\sigma_{\parallel}$,
with an $L^*$ elliptical having $\sigma_{\parallel} = 220$ km
s$^{-1}$ (\cite{fab76}), and following Turner \etal\ (1984) and
Gott (1977), we assign $\sigma = \sqrt{1.5}~\sigma_{\parallel}$.
In choosing $z_{lens}$, there is a trade-off between placing the
lens closer or farther from us.  Increasing $z_{lens}$ means the
inferred luminosity of B and, consequently, $\sigma$ will increase;
however, $D_{LS}$ will fall.  Conversely, moving the lens to lower
redshifts increases $D_{LS}$ but decreases the inferred luminosity
and $\sigma$ and consequently $\theta_{E}$.  For $z_{source} =
0.93$, the maximum of $\theta_E$ versus $z_{lens}$ is shallow, with
$(\theta_E)_{max} = 0\farcs56$ at $z_{lens}\approx0.35$.  At this
redshift, component B is 2 magnitudes fainter than $L^{*}$.  Note
the aperture correction and the choice of $M_B^*$ do not strongly
affect the calculation since $\theta_E \sim L^{1/2}$ through the
Faber-Jackson relation.

The predicted Einstein ring radius is nearly a factor of 4 smaller
than the observed separation between the assumed lensing galaxy
(component B) and resulting images (the nuclear structure of A);
equivalently, the mass-to-light ratio we infer for B is 4 times
larger than typical ellipticals. However, this lensing scenario
remains plausible.  The gravitational potential of B may be
elliptical in which case images will be formed inside and outside
the Einstein radius (\eg,
\cite{gro88}).  The potential could be elliptical even though its
$K$-band appearance is circular; a similar situation occurs in the
case of F10214 (\cite{bro95,eis96}).  In addition, the intrinsic
$\sim$~25\% scatter in $\sigma$ from the Faber-Jackson relation
(\cite{djo87}) could lead to a $\sim$~50\% increase in $\theta_E$,
and mass distributions more sharply peaked than the isothermal
sphere will also boost the expected separation.  Extinction is an
unlikely explanation since components A and B have comparable $JHK$
colors (\cite{soi94}), and the amount of dust needed to make
$\theta_{E} = 2\arcsec$ would be enormous ($A_{V}
\approx 18$).  

F15307 bears a strong resemblance to probable quadruple-image
gravitational lens systems, \eg, PG~1115+080
(\cite{kri93,you81,wey80}), MG~2016+112 (\cite{law84,gar94,sch86}),
and especially MG~J0414+0534 (\cite{ang94,ann93,hew92}). Therefore,
an alternative lensing scenario is one where components B and C and
the nuclear structure of A are all produced by lensing, with the
lensing galaxy producing the diffuse emission which lies inside and
around these sources.  If the $z=0.93$ source lies just inside of
the fold of the lensing potential's diamond caustic (\eg, Figure 6c
and 6d of
\cite{bla92}), four images will be produced with a 
configuration similar to the arrangement of B, C, and the nuclei of
A.  The source would need to be extended to explain the extended
central structure of A and the resolved nature of B. One objection
to this scenario is that none of the components, especially
component A, appear tangentially stretched, though such an effect
probably requires a factor of $\sim$ 2 better spatial resolution to
discount completely.  Moreover, the lensing galaxy must be
considerably underluminous.  For instance if the lensing galaxy is
located at the most probable location, $D_{OS} = 2 D_{LS}$ in a
flat universe (\cite{pee93,got89}) which corresponds to
$z_{lens}=0.4$, an $L_*$ elliptical has $K \approx 16$~mag and $r_{e}
\approx 2\arcsec$ (6\hone\ kpc at $z=0.4$).  Such a galaxy might account for
the $\sim$~6\arcsec\ diffuse emission around F15307; however, the
central region of such a galaxy would be apparent in our images so
if the F15307 is a four-image lens, the lensing galaxy is likely
underluminous for its mass.

Note that our identification of F15307-A as an elliptical galaxy
(\S~3.1.1) is incompatible with most lensing scenarios.  The galaxy
shows no sign of distortion from elliptical symmetry which would
occur if it was a lensed source at $z=0.93$, and the apparent
coincidence of the galaxy with the presumed lensed images (the
nuclear structure of A) rules out the galaxy being the foreground
lens since lensed images coinciding with their lenses are strongly
demagnified (\eg,
\cite{bla92}).  If in fact F15307 is a lensed system, F15307-A is
probably composed of one or two tangentially stretched images with
diffuse emission from the foreground lensing galaxy, leading us to
misidentify the outer 1\arcsec-3\arcsec\ emission as being centered
on the nuclear structure.  One final important point: even if the
system is lensed, the fact that the components are extended means
that, just like FSC 10214, the total magnification cannot be
enormous; the system must be an intrinsically luminous galaxy.

\subsection{Or a Cannibal Elliptical at z = 0.93?}

Alternatively, F15307 may be intrinsically the most luminous galaxy
in the known Universe.  Its physical characteristics and
environment are consistent with the idea that F15307 is a higher
redshift and more luminous analogue of the most luminous galaxies
in the local universe, the ultraluminous ($L_{IR} \gtrsim 10^{12}
L_{\sun}$) \IRAS\ galaxies (\cite{san88a}). Most ultraluminous IR
galaxies (ULIRGs) possess disturbed morphologies and two or more
close companions (\cite{san88a,san88b,arm90,mel90,lee94,clo95}); a
large fraction also possess close double nuclei with $\lesssim$~1
kpc separations (\cite{gra90,car90,maj93,arm94}).  Evidence for
interactions in these systems seems to correlate with their total
luminosity as well as the presence of an active nucleus
(\cite{san88a,san92}).  Indeed the merging process is believed to
play a central role in the ULIRG phenomenon, though the issue of
whether the high far-IR luminosity arises from dust-enshrouded
active nuclei and/or compact starburst remains controversial (\eg,
\cite{san88a,con91,kor92}).  Similar processes could energize
the extraordinary far-IR luminosity and active nucleus in F15307.

There exist some similarities between F15307 and radio galaxies.
Its $K$ magnitude is comparable to $z\approx1$ radio galaxies, and
its $R-K$ and $B-R$ colors (\cite{cut94}) compared to its $K$
magnitude lie between radio galaxies and radio quasars
(\cite{dun89}). Also, F15307's radio power is considerable;
assuming a power-law spectrum, $L_{\nu}
\sim \nu^{-\alpha}$, the observed 8.2~GHz flux density (\cite{cut94})
extrapolates to a restframe 1.49 GHz flux density of
$\ex{6.0}{23}~\htwo$ W~Hz\perone\ and $\ex{6.5}{24}~\htwo$
W~Hz\perone\ for 
$\alpha$ = 0 and 1, respectively.  Using a canonical $\alpha \sim
0.7$, F15307 would be classified as radio-loud ($L_{1.49 GHz} >
\ex{2.5}{24} \htwo$ W~Hz\perone\ [\cite{wol90}]), with its radio
power exceeding that of typical ultraluminous \IRAS\ galaxies by a
factor of 10 (\cite{con91}) and the most radio-luminous ordinary
galaxies by a factor of 100 (\cite{con90}).

Circumstantial evidence argues that the dust in F15307 is not
widespread.  Our upper limit on very broad \ha\ emission (\S~3.2)
implies (\ha/\mgii) $\lesssim$ 35 for the broad lines.  Using
the expected ratio for a unreddened quasar spectrum (\cite{ost89})
and an ordinary interstellar extinction law (\cite{mat90}), our
data imply $A_V \lesssim 3$ along the line of sight to the region
scattering light from the broad-line region.  Also, the ratio of
the \ha\ line to the H$\delta$ line (\cite{hin95})
does not differ much from the value expected for pure
recombination.  Finally, the fact that the $R-K$ color is quite
blue compared to typical ellipticals (see below) suggests dust is
not widespread.  The dust which generates the large far-IR
luminosity may have a compact spatial distribution.

Since its surface brightness profile is well-described by a
de~Vaucouleurs law, our data suggest F15307-A is an elliptical
galaxy.  The measured $K$-band (restframe $J$-band) SB profile
should be a good tracer of F15307-A's dynamical structure since
galaxian $J$-band emission arises from old mass-tracing stellar
populations, although contributions from recent star formation are
not entirely negligible (\cite{bru93}), and is relatively
insensitive to extinction.  The properties derived from the SB
profile (\S~3.1.1) are in accord with the identification of
F15307-A as an elliptical galaxy.  The 6\hone\ kpc half-light
radius of F15307-A lies in the upper range seen in local giant
elliptical galaxies (\cite{pah95,san90}).  Based on the
calculations in
\S~4.1, a non-evolving elliptical at $z = 0.93$ has $R - K \approx
6$~mag; using our $K$-band measurement alone would suggest F15307-A
has a restframe $B$-band $\mu_{e} \approx 23$ mag~arcsec\perone, a
value which is also in the range of local giant ellipticals
(\cite{san90}).  However, the integrated light of F15307 has $R-K
\approx 3$~mag (\cite{cut94}), much bluer than a typical elliptical,
which may signal on-going star formation in one or more of the
components.

The fact that F15307 has more than two components leaves our
interpretation of the physical situation ambiguous.  Mergers of two
disk galaxies is believed to result in an elliptical
(\cite{barhern92} and references therein).  This phenomenon is seen
in numerical simulations of galaxy collisions (\eg,
\cite{too72,bar92}) and in reality: optical and IR imaging have
found $r^{1/4}$ profiles in the centers of advanced disk-disk
mergers (\cite{sch82,wri90,sta91}), and velocity dispersion
measurements argue that merging systems will evolve into
ellipticals (\cite{lak86,doy94}).  Based on its SB profile,
F15307-A is clearly an elliptical galaxy, but if we are observing
its formation via disk-disk merger how do we account for the other
companions?  Multiple merger events are suspected to be responsible
for the formation of cD ellipticals (\eg,
\cite{hau78}),
which do show multiple nuclei (\eg, \cite{mat64}); however, cD's
are exclusively found in clusters (\cite{bee83,mor75}), and no
cluster is known to be associated with F15307.\footnote{The next
most luminous \IRAS\ galaxy after F15307, P09104+4109, with a
luminosity of $\sim 6 \times 10^{12}~\htwo\ L_{\sun}$, is a cD
galaxy in a $z=0.44$ cluster (\cite{kle88,soi96}).}  Moreover, with
$R_e = 6\hone$~kpc, F15307-A would be somewhat small for a typical cD
galaxy (\cite{oeg91}).  However, multiple merger events might occur
in the formation history of all ellipticals, not just cD's
(\cite{wei94}).  F15307-A itself may have formed at a higher
redshift, and we are observing its subsequent interaction with
components B and C at $z = 0.93$, whose outcome will be the
consumption by component A of its two companions.


\section{Conclusions}

We have presented the highest spatial and spectral resolution
near-IR observations to date of the $\sim~10^{13}~\htwo~L_{\sun}$
\IRAS\ galaxy FSC~15307+3252 located at $z = 0.93$, apparently
the most luminous known galaxy.  We find the following results:

\noindent 1. Deep $K$-band images with 0\farcs4 resolution 
reveal at least three components to the system.  The brightest,
component A, is elliptical with a compact nucleus.  Component B, is
resolved and apparently circular, and the faintest component, C,
has an irregular morphology.

\noindent 2. The $K$-band surface brightness profile of F15307-A is
well-described by a de~Vaucouleurs $r^{1/4}$ law with $\mu_{e}
= 20.1 \pm 0.3$~mag~arcsec\pertwo\ and $r_{e} = 1\farcs4 \pm
0\farcs2$ (6\hone~kpc) after correction for seeing effects.  Its
effective radius is comparable to local giant ellipticals. After
removal of the de~Vaucouleurs profile, the core of F15307-A shows a
compact nucleus with extended structure $\sim$~0\farcs5 to the
southwest, possibly a second nucleus.

\noindent 3. Our 1.1--1.4 \micron\ (restframe optical) spectrum with
a resolution of 330~km~s\perone\ shows strong emission lines of
\oi, \ha~+~\nii, and \sii\ with velocity widths typical of Seyfert~2
galaxies.  The line excitation is also consistent with such a
classification.  \ha\ also has a strong $\sim$~1900~km~s\perone\
component but lacks a very broad $\sim$~10$^{4}$~km~s\perone\
component unlike the polarized \mgii\ \lam2798 line (\cite{cut94}).
The line emission may be extended on scales of
$\sim$~1\arcsec\ (4\hone\ kpc).

\noindent 4. The morphology of F15307 is reminiscent of known 
gravitational lensed objects, particularly quadruply-imaged sources
such as MG~J0414+0534.  Quantitative arguments are inconclusive,
though if the system is lensed, the absence of an obvious
foreground lens imply the lensing galaxy is underluminous for its
mass.  The fact that the components are extended means even if the
system is lensed, the $z=0.93$ source must be an intrinsically
luminous galaxy.

\noindent 5. Alternatively, F15307 may be an interacting system
with an intrinsically large luminosity, similar to local
ultraluminous \IRAS\ galaxies.  Some indications exist that the
system is not heavily extincted.  The $r^{1/4}$ profile suggests
F15037-A is an elliptical galaxy. It may be in the process of
forming at $z=0.93$ or else it formed at $z > 0.93$ and we are now
observing its interaction/merger with components B and C.

Additional observations should determine the nature of F15307.
\HST\ imaging should be able to identify if the system is lensed
but is not essential; ground-based IR observations in excellent
seeing or high resolution radio imaging should also suffice.  Color
information will be useful --- if the system is lensed, the
foreground lensing galaxy, most likely an elliptical, should be
distinct from the multiple images of the background lensed Seyfert.
Narrow-band imaging centered on emission lines will be a good test
for lensing as will long slit spectroscopy to compare the spectra
of components A and B.  One additional observation which would
serve as an empirical test is to search for CO emission from F15307
since the only two definite detections of CO emission at high
redshift are from the lensed sources F10214 and the Cloverleaf
quasar (\cite{sol92,bar94}).  Regardless of which way the issue is
settled -- if the system is a lensed Seyfert galaxy or an
interacting elliptical --- the system will be worthy of further
scrutiny.

\acknowledgements

We are grateful to Tom Broadhurst, Luis Ho, Joe Shields, and Steve
Zepf for useful discussions and thank Arjun Dey, George Djorgovski,
and Hy Spinrad for their comments on a draft of this work.  The
W. M. Keck Observatory is a scientific partnership between the
University of California and the California Institute of
Technology, made possible by the generous gift of the W. M. Keck
Foundation and support of its president, Howard Keck.  It is a
pleasure to thank Barbara Schaeffer, Wendy Harrison, and Imke de
Pater for their help with these observations.  J.~R.~G. is
supported by a fellowship from the Packard Foundation and M.~C.~L.
by an NSF Graduate Student Fellowship.

\newpage



\begin{deluxetable}{ccrrc}
\tablewidth{4.5in}
\small
\tablenum{1}
\tablecaption{Identification and Photometry of Sources in the
F15307 Field}
\tablehead{
\colhead{Object} & 
\colhead{} &
\colhead{$\Delta\alpha$ (\arcsec)}   &
\colhead{$\Delta\delta$ (\arcsec)}   & 
\colhead{$K$ magnitude\tablenotemark{a}} 
} 
\startdata

1  &             &   --19.2 &    13.3 &    21.0  $\pm$   0.3 \nl
2  &             &   --15.2 &    23.1 &    22.5  $\pm$   1.0 \nl
3  &             &   --14.4 &    20.4 &    18.92 $\pm$  0.04 \nl
4  &             &   --12.8 &   --9.2 &    21.4  $\pm$   0.5 \nl
5  &             &   --12.1 &    21.3 &    22.8  $\pm$    1.1 \nl
6  &             &   --11.7 &    25.9 &    18.10 $\pm$  0.03 \nl
7  &             &    --9.0 &    30.4 &    17.81 $\pm$  0.03 \nl
8  &             &    --8.3 &     7.5 &    20.00 $\pm$  0.07 \nl
9  &             &    --6.9 &     2.0 &    20.37 $\pm$  0.10 \nl
10 &             &    --3.2 &  --11.8 &    18.85 $\pm$  0.02 \nl
11 &             &    --2.0 &    10.6 &    20.83 $\pm$   0.16 \nl
12 &             &    --1.0 &   --7.5 &    20.6  $\pm$   0.2 \nl
13 & F15307--A   &      0.0 &     0.0 &    16.59 $\pm$  0.02 \nl
14 & F15307-B    &      1.4 &   --1.4 &    18.40 $\pm$  0.02\tablenotemark{b} \nl
15 & F15307-C    &      1.5 &   --0.2 &    19.62 $\pm$  0.03\tablenotemark{b} \nl
16 & bright star &      2.8 &    19.5 &    15.89 $\pm$  0.02 \nl
17 &             &      4.6 &  --11.8 &    19.54 $\pm$  0.10 \nl
18 &             &      6.5 &   --9.4 &    19.53 $\pm$  0.11 \nl
19 &             &      8.4 &    26.8 &    17.51 $\pm$  0.02 \nl
20 &             &      9.6 &   --8.5 &    19.8  $\pm$   0.2 \nl
21 &             &     12.2 &   --1.0 &    20.14 $\pm$  0.09 \nl
22 &             &     13.3 &    23.0 &    20.20 $\pm$  0.10 \nl
23 &             &     13.5 &    14.1 &    21.0  $\pm$   0.2 \nl
24 &             &     14.5 &     8.6 &    20.8  $\pm$   0.2 \nl
25 &             &     17.5 &     6.6 &    21.2  $\pm$   0.3 \nl
26 &             &     17.9 &   --4.9 &    18.93 $\pm$  0.06 \nl
27 &             &     21.0 &    19.0 &    20.4  $\pm$   0.2 \nl
28 &             &     23.1 &    18.1 &    20.2  $\pm$   0.2 \nl
29 &             &     23.8 &   --6.6 &    17.94 $\pm$  0.04 \nl

\tablenotetext{a}{All photometry done in a 3\arcsec\ diameter
aperture except for the components of F15307 which used 2\arcsec\
(component A), 1\farcs5 (B), and 1\arcsec\ (C) diameter apertures.
Values include a small aperture correction derived from the bright
star.  Errors were calculated by combining in quadrature the noise
in the photometry aperture with errors in the zero point and sky
level determination (\S~2.1).}
\tablenotetext{b}{After removal of the best-fitting deVaucouleurs
profile for component A, aperture-corrected photometry for
components B and C gives 18.6 $\pm$ 0.1 (1\farcs5 diameter
aperture) and 20.1 $\pm$ 0.3 (1\arcsec\ aperture), respectively
(\S~3.1.2).}

\enddata
\end{deluxetable}

\clearpage


\begin{deluxetable}{llccccc}
\tablewidth{0pt}
\tablenum{2}
\tablecaption{F15307 Emission Line Measurements}
\label{spec}
\tablehead{
\colhead{} & 
\colhead{} &
\colhead{$\lambda$} &
\colhead{$z$} &
\colhead{FWHM\tablenotemark{a}} &
\colhead{Flux} &
\colhead{Restframe equiv.\ width} \\
\colhead{} &
\colhead{} &
\colhead{(\AA)} & 
\colhead{} &
\colhead{(km s\perone)} & 
\colhead{(10$^{-19}$ W m\pertwo)} & 
\colhead{(nm)} 
}

\startdata

\oi & & 6300 &  0.9288 $\pm$ 0.0009 &  900 $\pm$ 300 &  7 $\pm$ 3 & 5 $\pm$ 2  \nl
    & & 6364 &  0.9288 $\pm$ 0.0009 &  900 $\pm$ 300 &  2.5 $\pm$ 1.2 & 1.6 $\pm$ 0.8 \nl         
			              	                                                                
\sii\tablenotemark{b} & & 6716 &  0.9230 $\pm$ 0.0014 &  900 $\pm$ 600 &  2.2 $\pm$ 1.8 &  1.5 $\pm$ 1.3  \nl
    & & 6731 &  0.9230 $\pm$ 0.0014 &  900 $\pm$ 600 &  5 $\pm$ 4 &  3 $\pm$ 3  \nl
			              	                                                                
\nii\tablenotemark{c} & & 6548 &  0.9280  $\pm$ 0.0004  &  900 $\pm$ 300 &  3.3 $\pm$ 1.2 &  2.1 $\pm$ 0.8 \nl
    & & 6583 &  0.9280  $\pm$ 0.0004  &  900 $\pm$ 300 &  10 $\pm$ 4\phn &  6 $\pm$ 2  \nl
			              	                                                                
\ha & (narrow)\tablenotemark{c}  & 6563 &  0.9280 $\pm$ 0.0004 &  900 $\pm$ 300 &  11 $\pm$ 4\phn &  8 $\pm$ 3  \nl
    & (broad)   & 6563 &  0.9280 $\pm$ 0.0004 &  1910 $\pm$  50\phn\phn & 39.4 $\pm$ 1.3\phn &   25.8 $\pm$ 0.9\phn \nl

\tablenotetext{a}{uncorrected for instrumental broadening}
\tablenotetext{b}{assumes high density limit for the ratio of the
two lines}
\tablenotetext{c}{FWHM fixed to be the same as \oi\ (see \S~3.2)}

\enddata
\end{deluxetable}

\clearpage


\clearpage


\begin{figure} 
\vskip -5in
\plotone{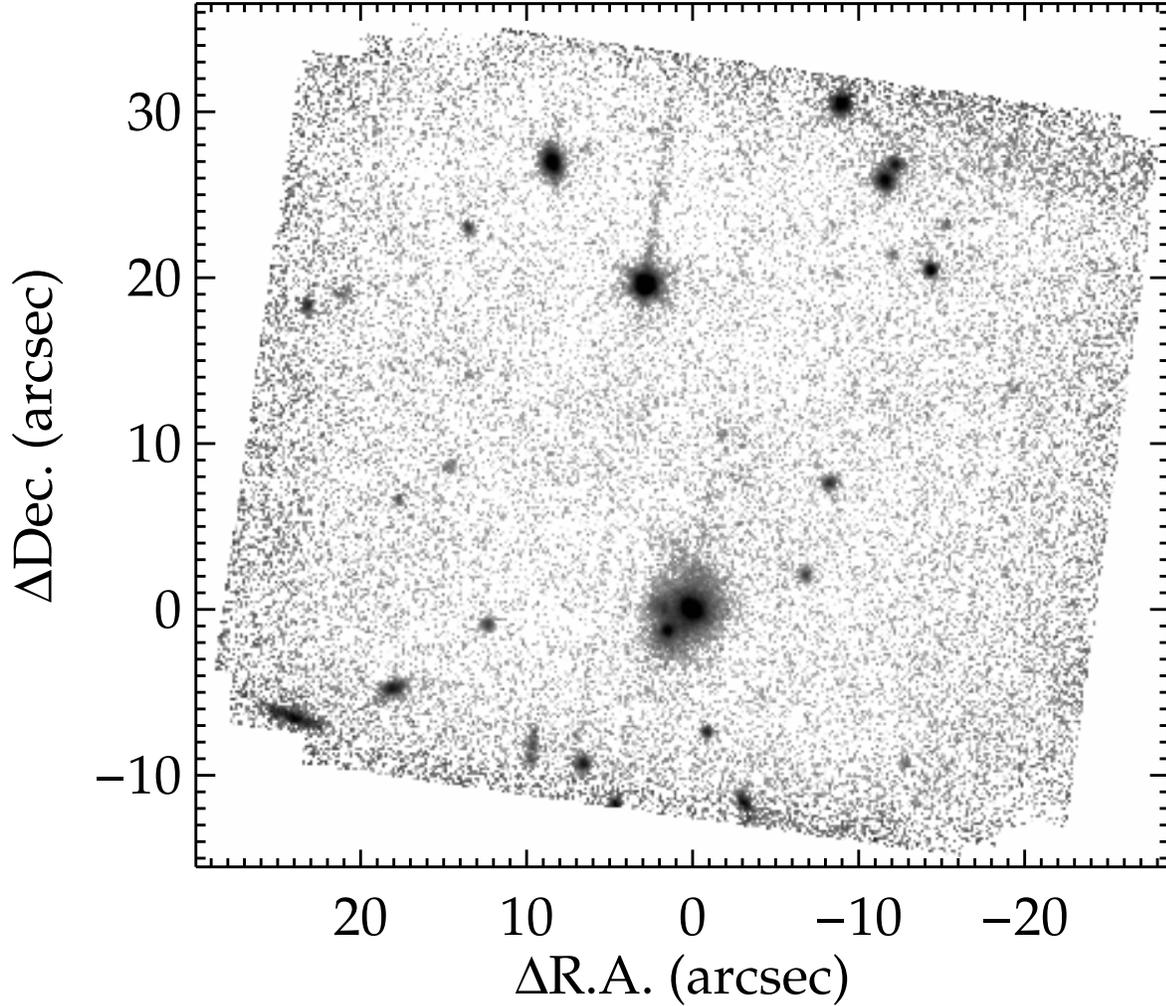}
\caption{$K$-band mosaic of the F15307 field obtained in 0\farcs4
seeing at the W. M. Keck Telescope. North is up and east to the
left. The greyscale is logarithmic. F15307 is the large galaxy with
multiple components at the origin of the axes.  The deepest portion
of the mosaic achieves 1~$\sigma$ noise of 23.9 mag~arcsec\pertwo,
and the faintest objects are $K \approx 21.5-22$. The long northern
tail of the brightest star in the field is an artifact of the data
acquisition and is not physically significant. For our assumed
cosmology ($q_{0} = 0.5$, $H_{0} = 100h$ km~s\perone~Mpc\perone),
1\arcsec\ corresponds to 4.2\hone~kpc.}
\end{figure}
\clearpage

\begin{figure}
\vskip -1.0 in
\plotone{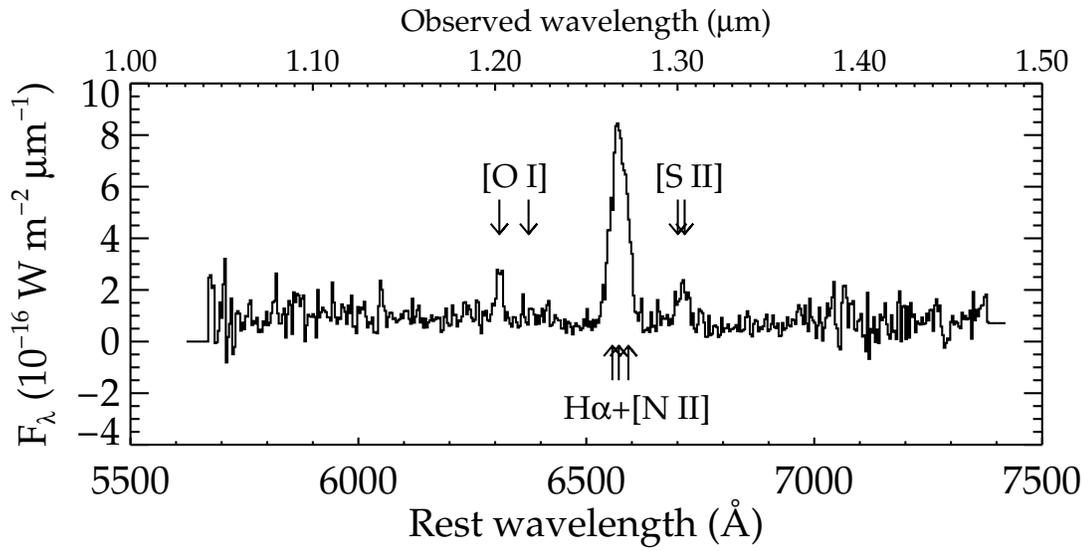}
\vskip -1.0in
\caption{$J$-band spectrum of F15307.  Emission lines of
[O I] \lam\lam6300, 6364; blended \ha~+~[N II] \lam\lam6548, 6583;
and [S II] \lam\lam6716, 6731 are observed and spectrally resolved.
The resolution is 330 km~s\perone\ (2 pixels).  The rest wavelength
scale assumes a redshift of 0.926.}
\end{figure}
\clearpage

\begin{figure}
\vskip -3in
\plotone{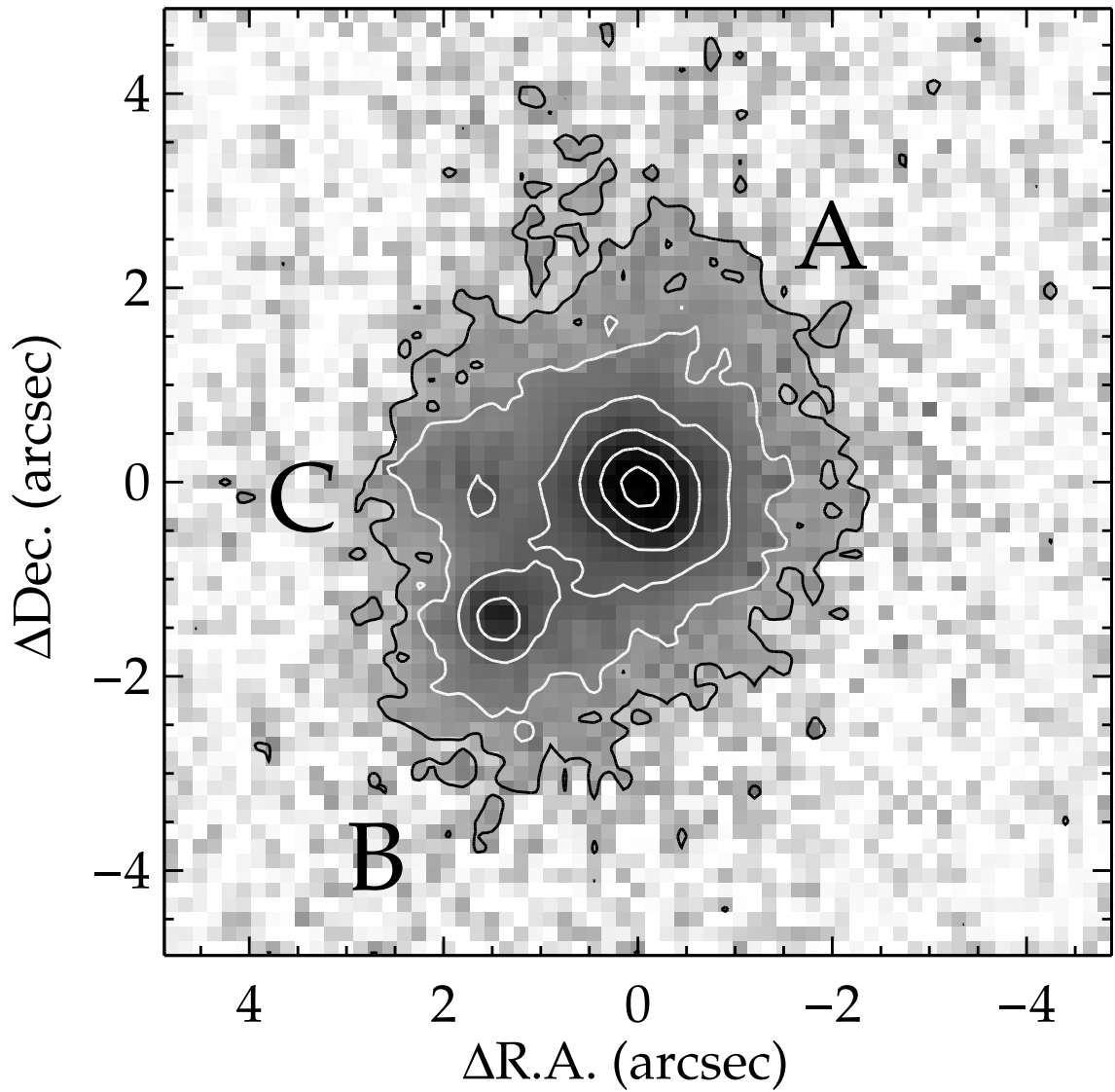}
\bigskip
\caption{A close view of the $K$-band image of F15307 along with
the designations we assign to the separate components 
based on their relative $K$ magnitudes (\S~3.1).  Again, the greyscale is
logarithmic. All three components are extended. Contours are
spaced by 1 magnitude (factor 2.5) with the brightest
contour being 16.3~mag~arcsec\pertwo.  For our assumed cosmology
($q_{0} = 0.5$, $H_{0} = 100h$ km~s\perone~Mpc\perone),
1\arcsec\ corresponds to 4.2\hone~kpc.}
\end{figure}
\clearpage

\begin{figure}
\plotone{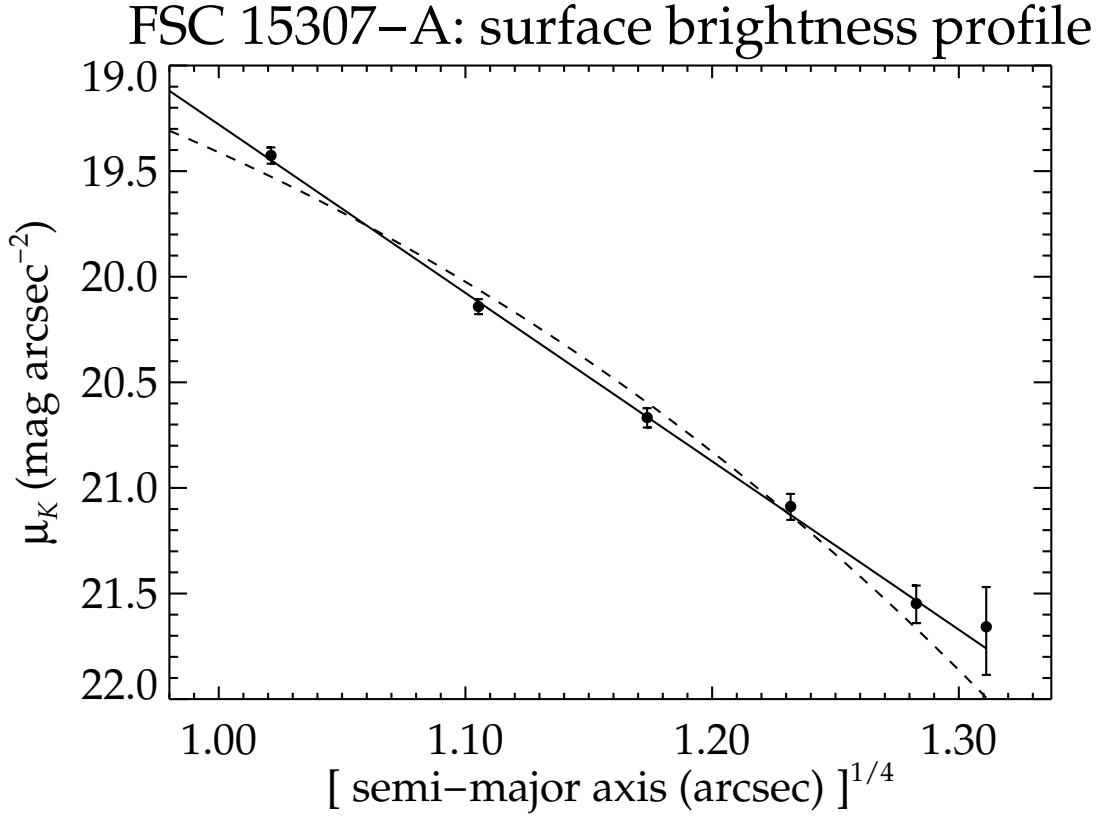}
\caption{A plot of the surface brightness profile of component A,
extracted using elliptical apertures of a fixed position angle and
axial ratio spaced by the 0\farcs4 seeing FWHM (\S~3.1.1). The solid line
shows a de~Vaucouleurs profile fitted from 1\arcsec--3\arcsec\
($\tilde{\chi}^2 = 0.4$), and the dashed line is an exponential
disk fit ($\tilde{\chi}^2 = 4.7$) --- clearly the de~Vaucouleurs fit
is superior.  After correcting for systematic effects due to
seeing, we find $\mu_e = 20.1 \pm 0.3$ mag~arcsec\pertwo\ and $r_e
= 1\farcs4 \pm 0\farcs2$ ($6\hone$~kpc for $H_{0} = 100h$
km~s\perone~Mpc\perone\ and $q_{0} = 0.5$), a size comparable to
local giant elliptical galaxies (\S~4.2).}
\end{figure}
\clearpage

\begin{figure}
\vskip -2in
\plotone{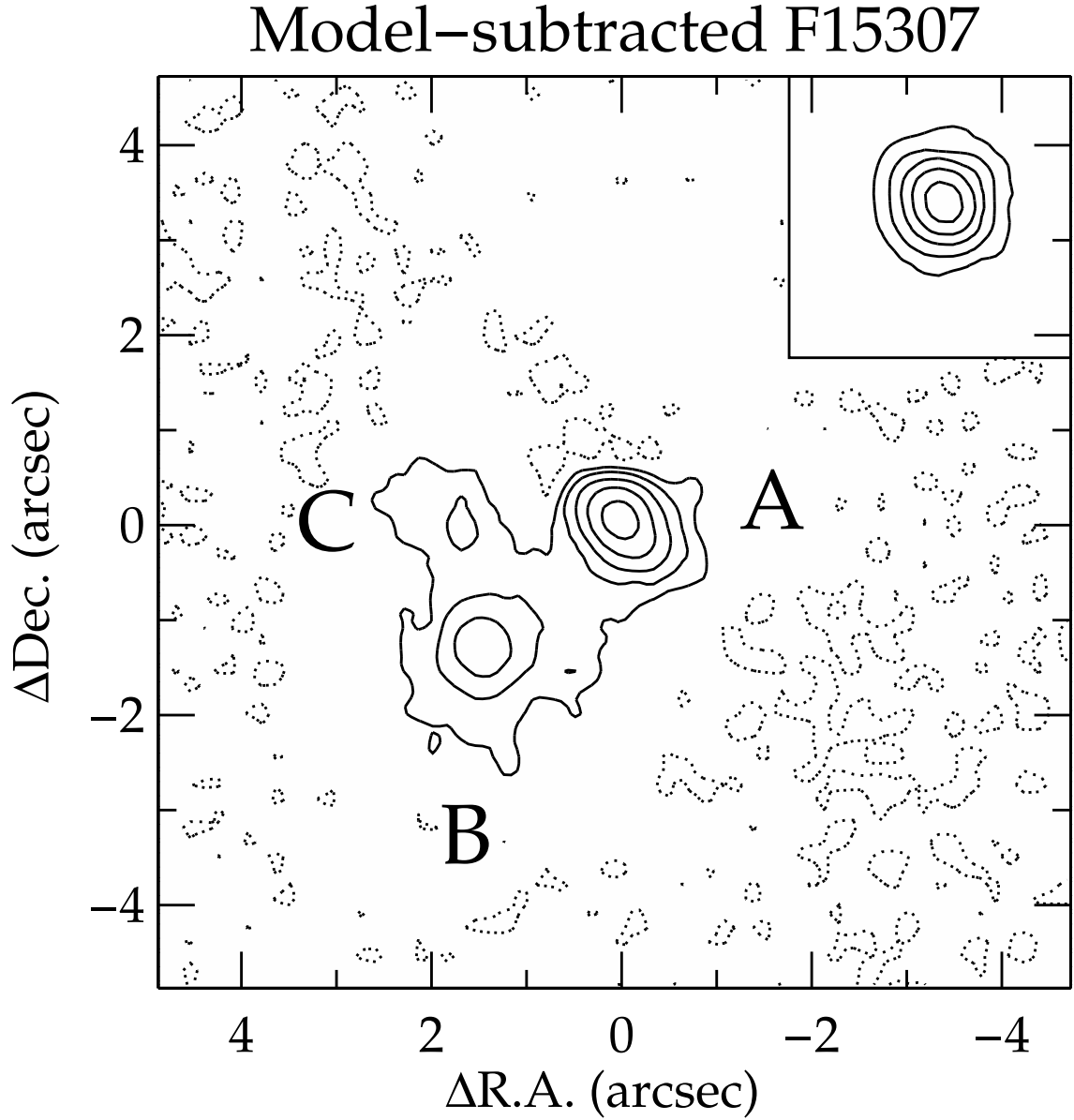}
\bigskip
\caption{$K$-band image of F15307 after removal of the
best-fitting de~Vaucouleurs profile centered on component A (\S
3.1.1).  The inset shows the point spread function as measured by
the bright star in the field for comparison.  The contours for both
images start at 64\% of the peak value and decrease by factors of
2.5.  The dotted contours represent --1\% of the peak.  The
residuals at the center of component A show an unresolved source
with an extension or second nucleus $\sim$~0\farcs5 to the
southwest.  Components B and C are clearly seen as is a bridge of
emission between A and B.}
\end{figure}
\clearpage

\end{document}